\documentstyle[aps,preprint,prd]{revtex}

\begin{document}
%
%

\draft

\title{Unquenching the $\rho$ meson}

\author{Derek B. Leinweber\cite{newaddr} and Thomas D. Cohen}

\address{
Department of Physics and Center for Theoretical Physics \\
University of Maryland, College Park, MD 20742
}

\date{July 9, 1993}

\maketitle

\begin{abstract}
Two-pion induced self-energy contributions to the $\rho$-meson mass
are examined in relation to the quenched approximation of QCD, where
the physics associated with two-pion intermediate states has been
excluded from vector-isovector correlation functions.  Corrections to
quenched QCD calculations of the $\rho$-meson mass are estimated to be
small at the order of a few percent of the $\rho$-meson mass.  The
two-pion contributions display nonanalytic behavior as a function of
the pion mass as the two-pion cut is encountered.  The implications of
this nonanalytic behavior in extrapolations of full QCD calculations
are also discussed.  We note that for full QCD, the error made in
making a linear extrapolation of the $\rho$ mass, neglecting
nonanalytic behavior, increases as one approaches the two-pion cut.
\end{abstract}

\pacs{12.38.-t, 12.38.Gc, 14.40.-n, 14.40.Cs}

\narrowtext

\section{INTRODUCTION}

   The lattice regularized approach to quantum field theory provides
the best forum for the examination of the fundamental nonperturbative
aspects of QCD.  In the low momentum transfer regime, it is the only
approach which in the foreseeable future holds a reasonable promise of
confirming or rejecting the validity of QCD as the underlying theory
of the strong interactions.

   Recently, it has become possible to perform quenched QCD
calculations in which all systematic uncertainties are quantitatively
estimated.  Thus if the effects of quenching can be understood, the
validity of QCD may be tested in the nonperturbative regime.  Of
particular note is the recent determination of the QCD coupling
constant, $\alpha_{\overline{MS}}$, from the $1S-1P$ mass splitting of
charmonium \cite{elkhadra92}.  In this case one believes that the
effects of quenching may be estimated with minimal model dependence.
Corrections to the utilization of the quenched approximation of QCD
have been estimated and are currently the dominant source of
uncertainty in the final predictions
\cite{paramspace}.

   In this paper we will continue efforts along this line through the
examination of systematic uncertainties in hadron mass spectrum
calculations.  In particular, the importance of the two-pion induced
self-energy contribution to the $\rho$-meson mass is evaluated in
relation to the quenched approximation of QCD and to full QCD.  In the
quenched approximation, the physics associated with two-pion
intermediate states has been excluded in the numerical simulations.

   This investigation is motivated by recent results from the GF11
group \cite{butler93} for the low-lying hadron mass spectrum in the
quenched approximation of QCD.  Their analysis is the first to
systematically extrapolate QCD calculations to physical quark mass,
zero lattice spacing and infinite volume.  Their predictions display
an impressive agreement with experiment.  Of eight hadron mass ratios,
six agree within one standard deviation and the remaining two ratios
agree within $1.6\sigma$.  However, since the quenched approximation
leaves out so much important physics, one might question whether these
results are actually {\it too} good \cite{sharpe92t}.

   In the perturbative regime, many of the effects of not including
disconnected quark loops when preparing an ensemble of gauge
configurations may be accounted for in a simple renormalization of the
strong coupling constant.  However, one also anticipates
nonperturbative effects in making the quenched approximation.  Unlike
a global renormalization of the coupling constant, these effects are
expected to be channel specific.  For example, the quenched
approximation of QCD leaves out the physics associated with the decay
of the $\rho$ meson to two pions.  This physics must be accounted for
and added to the quenched results prior to comparing with experimental
data.  Moreover, the calculated hadron masses are extrapolated as a
function of the pion mass squared to the point at which the pion mass
vanishes using linear extrapolation functions.  Such an approach
neglects nonlinear and indeed, nonanalytic behavior in the continuum
extrapolation function.  For example, in the case of the $\rho$-meson
mass, one expects nonanalyticity associated with the onset of the
two-pion cut.

   {\it A priori} one does not know the relative importance of
two-pion intermediate states of the $\rho$ meson in describing the
$\rho$ mass.  The substantial width of the $\rho$ meson at $151.5 \pm
1.2$ MeV indicates its coupling to pions is not small and
correspondingly these dynamics may have significant influence on the
$\rho$-meson mass.

   Geiger and Isgur were the first to study the possible importance of
the two-pion induced self-energy of the $\rho$ meson in relation to
lattice QCD calculations \cite{geiger90}.  Their results are based on
a string breaking quark model and predict large corrections to smooth
extrapolations of the $\rho$-meson mass at approximately 70 MeV.
These authors were motivated by the long standing problem of QCD
predictions for the $N/\rho$-mass ratio being too large.  Their hope
was the two-pion induced self-energy correction would sufficiently
raise the $\rho$ mass to solve this problem.  However, we now
understand that both the finite lattice spacing and the finite volume
of the lattice act together to push up this mass ratio.  Current
estimates \cite{butler93} for this ratio corrected to the infinite
volume, continuum limit are $1.22 \pm 0.11$ in excellent agreement
with the experimental value of 1.22.

   Geiger and Isgur \cite{geiger90} advocate using the nonlinearities
in the $\rho$ mass as a function of the quark mass to correct for the
linear extrapolation of lattice results.  We wish to stress that such
a procedure is only sensible for the extrapolation of full QCD
calculations.  As we shall argue, the entire two-pion induced
self-energy is absent in the quenched approximation and should be
added on to quenched QCD results prior to comparison with experiment.
We note that for the model of Ref.\ \cite{geiger90}, quenched lattice
calculations of the $\rho$-meson mass would be {\it reduced} by 160
MeV instead of increased by 70 MeV.  This would further exacerbate the
``$m_N/m_\rho$ problem'' discussed in their paper, rather that curing
it.

   In the quenched approximation, the $\rho$ meson cannot decay to two
light pseudoscalars.  As discussed in Ref.\ \cite{cohen92a} it is not
possible to generate intermediate states of the $\rho$ meson in the
quenched approximation in which one has two isovector pseudoscalars.
One might worry about the presence of the light isoscalar pseudoscalar
$\eta'$ which fails to obtain its heavy mass in the quenched
approximation \cite{cohen92a,bernard92,sharpe92,lighteta}.  However,
decay of the $\rho$ meson to a $\pi$ $\eta'$ is forbidden by G-parity
and decay to two $\eta'$-mesons is of course forbidden by the isospin
invariance of the strong interactions.  Hence the physics associated
with two-pion intermediate states of the $\rho$ should simply be added
onto the results extracted from calculations of quenched QCD, provided
the lattice spacing is defined by physical observables which do not
have a similar dependence on pion decays.

   Current calculations of full QCD typically employ quark
masses which place $2 m_\pi > m_\rho$.  As a result, the functional
form of the extrapolation function should account for the nonanalytic
behavior in the $\rho$-meson mass as the two-pion cut is encountered.
On the lattice, the spectral density does not have a cut but rather a
series of poles at the points satisfying
\begin{equation}
\sqrt{s} = 2 \, \left (m_\pi^2 + p_n^2 \right )^{1/2} \, ,
\end{equation}
where $p_n$ are the discrete momenta allowed on the lattice.
Obviously, to fully account for the two-pion induced self-energy of
the $\rho$, one must first extrapolate the lattice results to zero
lattice spacing and infinite volume prior to extrapolating the quark
masses to physical values.

   In evaluating the integrals describing the coupling of pions to the
$\rho$ one must take into account the $q^2$ dependence of the $\rho$
to two $\pi$ coupling constant $g_{\rho \pi \pi}$ reflecting the
internal structure of these mesons.  While this $q^2$ dependence was
extracted from a string breaking quark model in Ref.\ \cite{geiger90},
we elect to take a more agnostic approach and consider different
methods of cutting off the integral.  In particular we investigate a
sharp $\theta$-function cutoff and a dipole cutoff.  The underlying
reason for our agnosticism is the belief that models of the sort
underlying Ref.\ \cite{geiger90} are not likely to correctly describe
the structure of pseudo-Goldstone bosons such as the pion, as they do
not incorporate chiral symmetry.  While there are models of the
$\rho\pi\pi$ vertex which incorporate chiral symmetry and chiral
symmetry breaking \cite{praschifka87,roberts89,hollenberg92}, these
models are based on particular dynamical assumptions.  Accordingly, it
is difficult to assess the reliability of such models.  Instead, we
consider a range of possibilities for the vertex and to simplify this
task, we consider convenient phenomenological forms.

   One other paper addressing this issue \cite{degrand91} sidesteps
the problems surrounding the $q^2$ dependence of $g_{\rho \pi \pi}$ by
fixing $g_{\rho \pi \pi}$ to a constant and making two subtractions of
the divergent integral at $q^2 = 0$.  These subtractions are absorbed
into a mass and wave function renormalization.  However, this approach
excludes any analysis of $\rho$-meson mass extrapolations as the
subtraction terms themselves have an unknown $m_\pi$ dependence which
has been lost in the renormalization procedure.  Moreover,
contributions from virtual two-pion states have been absorbed into the
bare lattice parameters which is inconsistent with the dynamics
contained in the quenched approximation.

   The outline of this paper is as follows.  In Section II the model
used in examining the two-pion induced self-energy is outlined.  Two
methods for regulating the divergent self-energy are explored.  In
section III the relevance of the self-energy corrections to quenched
QCD simulations is discussed.  Section IV addresses the quark mass
extrapolation of full QCD calculations and the importance of nonlinear
behavior in the $\rho$-meson mass.  Finally, the implications of this
investigation are summarized in Section V.

\section{THE SELF-ENERGY}

   In modeling the two-pion induced self-energy of the $\rho$ meson,
$\Sigma_{\rho\pi\pi}$, the standard $\rho\pi\pi$ interaction motivated
by low-energy current algebra is used.  The effective Lagrange
interaction has the form \cite{herrmann92}
\begin{equation}
{\cal L}_{\rm int} = -i \, g_{\rho\pi\pi} \, \rho^\mu \left ( \pi
\stackrel{\textstyle \leftrightarrow}{\partial_\mu} \pi \right )
+ g_{\rho\pi\pi}^2 \, \pi^2 \, \rho^2
\, .
\end{equation}
The pions are further assumed to interact exclusively through the
$\rho$ channel as summarized in the following Schwinger-Dyson equation
for the $\rho$ propagator
\begin{equation}
G_{\mu\nu} = G_{\mu\nu}^0 + G_{\mu\sigma}^0 \, \Sigma^{\sigma\tau} \,
G_{\tau\nu} \, ,
\end{equation}
where
\begin{equation}
G_{\mu\nu}^0 = {-i \over q^2 - M_0^2 + i \epsilon}\, \left (
g_{\mu\nu} - {q_\mu q_\nu \over q^2}  \right ) \, ,
\end{equation}
in Landau gauge, and $M_0$ is the bare $\rho$-meson mass.  The
self-energy $\Sigma_{\rho\pi\pi}$ is defined through the solution
\begin{equation}
G_{\mu\nu} = {-i \over q^2 - M_0^2 - \Sigma_{\rho\pi\pi} + i
\epsilon}\, \left ( g_{\mu\nu} - {q_\mu q_\nu \over q^2} \right ) \, ,
\end{equation}
where
\begin{equation}
\Sigma^{\sigma\tau} \equiv \Sigma_{\rho\pi\pi} \left (
g^{\sigma\tau} - {q^\sigma q^\tau \over q^2}  \right ) \, .
\end{equation}
$\Sigma^{\sigma \tau}$ is given by the standard one loop integrals
\begin{equation}
- i \, \Sigma^{\sigma\tau} = \int
{d^4 k \over \left ( 2 \, \pi \right )^4} \,
g_{\rho\pi\pi}^2 \, \left \{
{\left ( q^\sigma - 2\, k^\sigma \right )
 \left ( q^\tau - 2\, k^\tau \right )
\over
\Bigl [ \left ( q - k \right )^2 - m_\pi^2 + i \epsilon \Bigr ]
\Bigl [ k^2 - m_\pi^2 + i \epsilon \Bigr ] }
- { 2 g_{\mu \nu} \over k^2 - m_\pi^2 + i \epsilon } \right \}
\, .
\label{oneloop}
\end{equation}
Physically, the integral is convergent due to the momentum dependence
of $g_{\rho\pi\pi}$.  However, this momentum dependence is unknown.
In light of this uncertainty, it is reasonable to parameterize the
momentum dependence in terms of some given functional form with an
adjustable parameter controlling how the function falls off as a
function of momentum transfer.  We consider two regulation
prescriptions.  In one, we assume a monopole form for
$g_{\rho\pi\pi}$, and for comparison, we also consider a sharp
$\theta$-function cutoff.

   The simplest fashion for introducing a covariant cutoff function
is through the use of a dispersion relation.  The second term of
(\ref{oneloop}) is $q$ independent and serves only to subtract the
quadratic divergence of the first term in maintaining current
conservation.  As a result, we write a dispersion relation for
$\Sigma(q^2)$ with one subtraction at $q^2 = 0$,
\begin{equation}
\Sigma(q^2) \equiv {1 \over \pi}
\int_0^\infty ds \, {q^2 \over s} \,
{ {\rm Im}\ \Sigma(s) \over s - q^2 }
\, .
\label{disp}
\end{equation}
Of course, the imaginary part of $\Sigma_{\rho\pi\pi}$ may be easily
determined using any number of techniques for rendering the integral
of (\ref{oneloop}) finite.  The imaginary part is
\begin{equation}
{\rm Im}\, \Sigma_{\rho\pi\pi}(q^2) = {g_{\rho\pi\pi}^2 \over 48 \pi}
\,
q^2 \,
\left ( 1 - {4 \, m_\pi^2 \over q^2} \right )^{3/2} \,
\theta \left ( q^2 - 4 \, m_\pi^2 \right ) \, .
\label{Imsigma}
\end{equation}
The value of $g_{\rho\pi\pi}$ at $q^2 = m_\rho^2$ is fixed by equating
the imaginary parts of
\begin{equation}
M_0^2 + \Sigma_{\rho\pi\pi} \equiv \left ( m_\rho + {i \, \Gamma \over
2 } \right )^2 \, ,
\end{equation}
at $q^2 = m_\rho^2$.  The physical values \cite{pdg90} $m_\rho =
768.1$ MeV and $\Gamma = 151.5$ MeV fix $g_{\rho\pi\pi}$ at $\sim
6.0$.

\subsection{\pmb{$\theta$}-Function Cutoff}

   To illustrate the physics associated with the real part of the
self-energy we first consider the integral of (\ref{disp}) cut off
covariantly by a sharp $\theta$-function at $s = \Lambda^2$.  The
functional form is
\begin{equation}
{\rm Re}\, \Sigma_{\rho\pi\pi} = {g_{\rho\pi\pi}^2 \over 48 \, \pi^2}
\, q^2 \,
\Biggl \{
\ln \left ( { 1 - \sigma_\Lambda \over 1 + \sigma_\Lambda} \right )
+ {8 \, m_\pi^2 \over q^2} \, \sigma_\Lambda
 - \sigma_q^3 \, \ln \left (
     {\sigma_\Lambda - \sigma_q \over \sigma_\Lambda + \sigma_q }
                    \right) \; \Biggr \} \, ,
\label{sigmacut}
\end{equation}
where
%
\begin{equation}
\sigma_q = \left ( 1 - {4 \, m_\pi^2 \over q^2} \right )^{1/2}
\, {,\ \rm and\ \  }
\sigma_\Lambda = \left ( 1 - {4\, m_\pi^2 \over \Lambda^2} \right
)^{1/2} \, .
\end{equation}

   Figure \ref{thetacut} illustrates the real part of
$\Sigma_{\rho\pi\pi}$ evaluated at $q^2 = m_\rho^2$ for a variety of
cutoffs ranging from slightly above $m_\rho^2$ to 4
GeV${}^2$.  For small cutoffs, most of the strength in the integral
lies below the $\rho$ mass and consequently the $\rho$ mass is pushed
up due to the mixing with pion states.  Of course, this behavior is
completely consistent with that anticipated by simple quantum
mechanical arguments.  For larger cutoffs the strength above the
$\rho$ mass acts to reduce the $\rho$ mass.

\subsection{Dipole Cutoff}

\begin{figure}
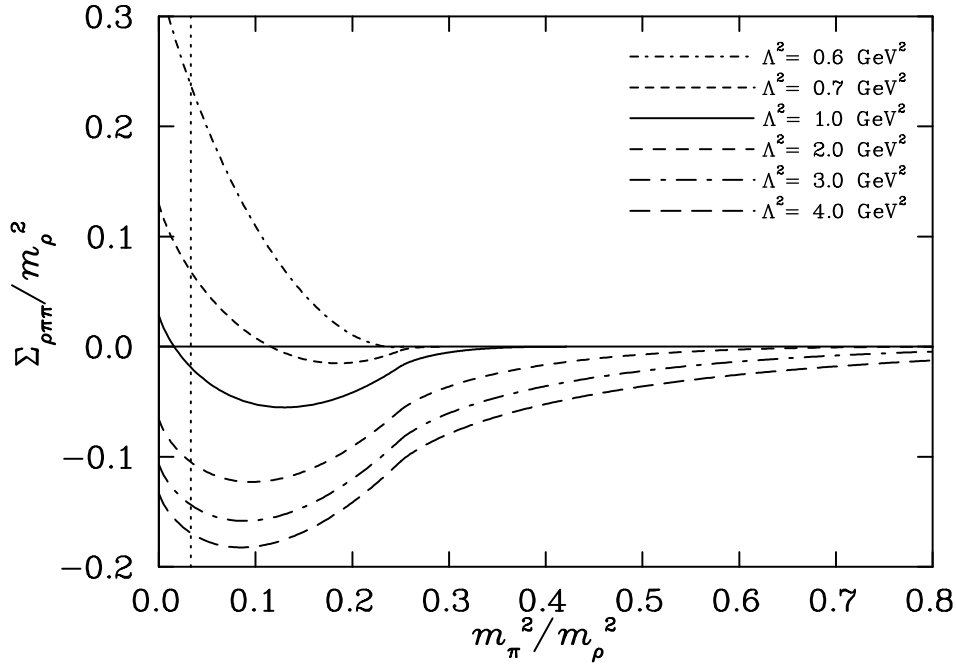

\caption{
The real part of $\Sigma_{\rho\pi\pi}$ from (\protect\ref{sigmacut})
evaluated at $q^2 = m_\rho^2$ for a variety of cutoffs, at $s =
\Lambda^2$.  In this and the following figures, the finely dashed
vertical line marks the position of the physical point.  For small
cutoffs, most of the strength in the integral lies below the $\rho$
mass and as a result the $\rho$ mass is pushed up.  }
\label{thetacut}
\end{figure}

   While the $\theta$-function is useful as an illustrative tool, it
suffers from being physically artificial and the results can be very
sensitive to the value of the cutoff, as illustrated in figure
\ref{thetacut}.  In an attempt to better represent the $q^2$
dependence of $g_{\rho\pi\pi}$ a monopole form for each vertex is
introduced and the dispersion relation of (\ref{disp}) is evaluated
with
\begin{equation}
g_{\rho\pi\pi}^2 \to g_{\rho\pi\pi}^2
\left ( q^2 + \Lambda^2 \over s + \Lambda^2 \right )^2 \, .
\label{gqdep}
\end{equation}
This form maintains the normalization of $g_{\rho\pi\pi}$ defined at
the physical $\rho$ mass and renders the integral of (\ref{disp})
finite.

   Of course, this approach is not without a few unphysical side
effects.  The most obvious problem is the introduction of spurious
poles in the space like region for the $s$ dependence of
$g_{\rho\pi\pi}$.  However, the dispersion integral only samples the
time like region and the presence of these unphysical poles should not
affect the results.  One could consider other functional forms.
However, the effective physical value for the regulator mass,
$\Lambda$, is itself unknown.  Our aim is to estimate the importance
of the two-pion induced self-energy relative to the $\rho$ mass, as
opposed to attempting to evaluate the actual correction.  For this
reason we view a consideration of the dipole regulator to be
adequate.

   Evaluation of the dispersion relation of (\ref{disp}) with
(\ref{gqdep}) leads to the following functional form for the real part
of $\Sigma_{\rho\pi\pi}$
\begin{eqnarray}
{\rm Re \ }\Sigma_{\rho\pi\pi} = {g_{\rho\pi\pi}^2 \over 48 \, \pi^2}
\, q^2 \,
\Biggl \{ &&
\left ( 1 + {8 \, m_\pi^2 \over q^2} + {12 \, m_\pi^2 \over \Lambda^2}
\right )
\left ( 1 + {q^2 \over \Lambda^2} \right )
\nonumber \\
&& + \left ( 1 + { 10 \, m_\pi^2 \over \Lambda^2} +
            {6 \, m_\pi^2 \, q^2 \over \Lambda^4} \right )
     \beta_\Lambda \, \ln \left ( {\beta_\Lambda - 1 \over
                                   \beta_\Lambda + 1 } \right )
\label{sigmadip} \\
&& - \sigma_q^3 \ln \left (
     {1 - \sigma_q \over 1 + \sigma_q} \right) \; \Biggr \}  \, ,
\nonumber
\end{eqnarray}
where
%
\begin{equation}
\sigma_q = \left ( 1 - {4 \, m_\pi^2 \over q^2} \right )^{1/2}
\, {,\ \rm and\ \  }
\beta_\Lambda = \left ( 1 + {4\, m_\pi^2 \over \Lambda^2} \right
)^{1/2} \,
{}.
\end{equation}
%
The imaginary part is recovered as in (\ref{Imsigma}).

\begin{figure}
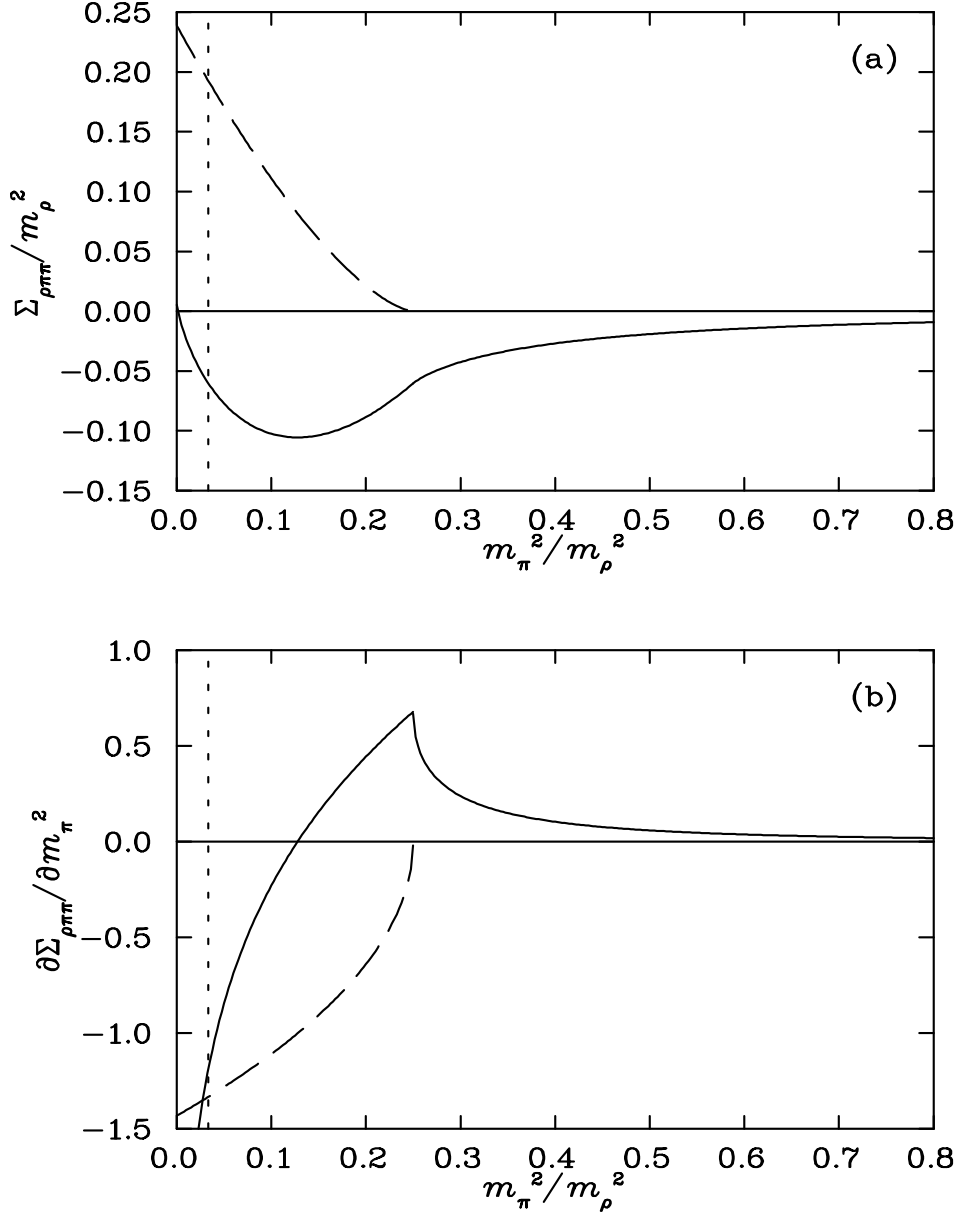

\caption{
The two-pion induced self-energy (a) and its derivative with respect
to $m_\pi^2$ (b) for a regulator mass of $\Lambda^2 = 2$ GeV${}^2$.
Both real (solid line) and imaginary (dashed line) parts are
illustrated.  The derivative clearly displays the nonanalytic behavior
encountered at $m_\rho = 2 m_\pi$.  }
\label{DRhoMass2GeV}
\end{figure}

   Figure \ref{DRhoMass2GeV} illustrates the real part of the
self-energy and its derivative with respect to $m_\pi^2$ at $\Lambda^2
= 2$ GeV${}^2$.  The derivative clearly displays the nonanalytic
behavior encountered at $m_\rho = 2 m_\pi$.  The second derivative is
discontinuous at $m_\rho = 2 m_\pi$ and is infinite from above.  The
imaginary part of the self-energy is also illustrated in figure
\ref{DRhoMass2GeV}.

\begin{figure}
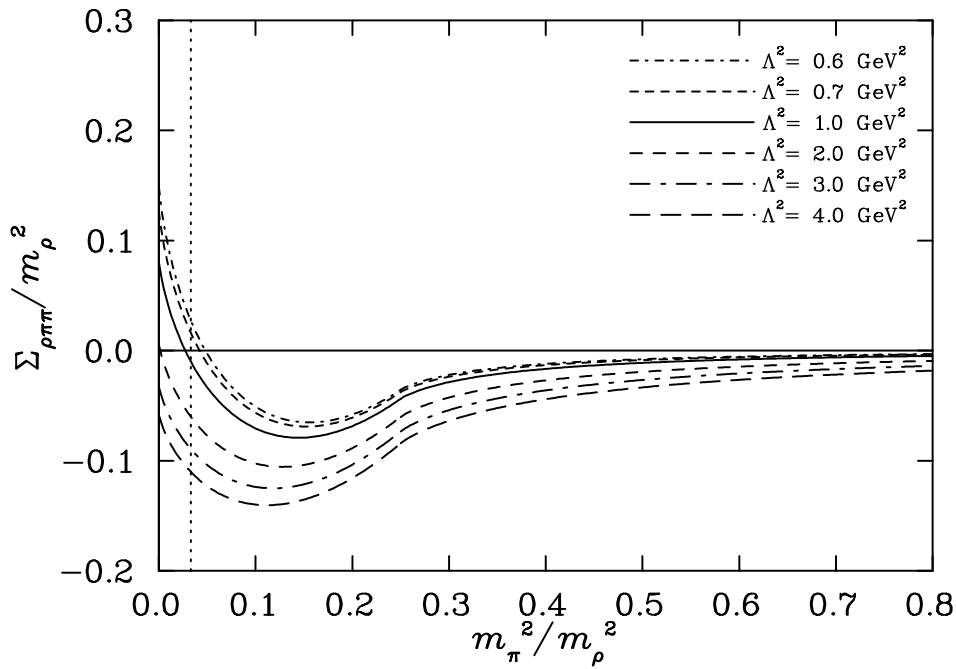

\caption{
The real part of the self-energy for dipole regulator masses $\Lambda$
taking the same values used in figure \protect\ref{thetacut}.  The
sensitivity of the results to the value of $\Lambda$ is greatly
reduced.  }
\label{DRhoMass}
\end{figure}

   A comparison with figure \ref{thetacut} indicates that at
$\Lambda^2 = 2$ GeV${}^2$ the results are not too sensitive to the
manner in which the integral is regulated.  Figure \ref{DRhoMass}
illustrates the real part of the self-energy for the same values of
$\Lambda^2$ used in figure \ref{thetacut}.  The sensitivity of the
results to the value of $\Lambda$ is greatly reduced and all curves
display the same qualitative behavior.

   In the limit of $\Lambda \to \infty$ both (\ref{sigmacut}) and
(\ref{sigmadip}) reduce to
\begin{eqnarray}
{\rm Re\ }\Sigma_{\rho\pi\pi} = {g_{\rho\pi\pi}^2 \over 48 \, \pi^2}
\, q^2 \,
\Biggl \{ &&
1 + {8 \, m_\pi^2 \over q^2} +
\ln\left ( {m_\pi^2 \over q^2} \right ) -
\sigma_q^3 \ln \left (
               {1 - \sigma_q \over 1 + \sigma_q } \right) \;
\nonumber \\
&& - \ln\left ( {\Lambda^2 \over q^2} \right )
\Biggr \} \, , \label{limlambda}
\end{eqnarray}
displaying the logarithmic divergence as $\Lambda^2 \to \infty$.

\section{APPLICATION TO QUENCHED QCD}

   The effects of quenching QCD may be categorized as perturbative or
global effects and nonperturbative or channel specific effects.  As
discussed in the introduction, the effects in the perturbative regime
may be accounted for through a simple renormalization of the coupling
constant.  However, the effects in the nonperturbative regime will, of
course, be channel specific.  The $\rho$-meson channel might be
particularly vulnerable to the nonperturbative effects of quenching
QCD due to the fact that the $\rho$ is unstable in full QCD and stable
in the quenched approximation.  As we shall see however, our estimated
``unquenching'' corrections turn out to be rather small.

   Figure \ref{gf11} displays $\rho$ and squared pion masses for the
three lightest quark masses used in the quenched QCD analysis of the
GF11 group \cite{butler93} for the $32 \times 30 \times 32 \times 40$
lattice at $\beta = 6.17$ for 219 configurations.  The dashed line
illustrates the linear relationship assumed in extrapolating the
hadron masses to the critical point where the pion mass vanishes.

   Provided the lattice spacing is defined by physical observables
which do not have a similar dependence on pion decays, the physics
associated with two-pion intermediate states of the $\rho$ should
simply be added onto the results extracted from calculations of
quenched QCD.  The correction is particular to the $\rho$ meson and is
unlikely to be accounted for if the lattice spacing is fixed by the
nucleon mass for example.

The solid line of figure \ref{gf11} illustrates the addition of the
self-energy correction to the linear extrapolation where the regulator
mass $\Lambda^2 = 1$ GeV${}^2$ has been selected \cite{corrlatt}.
This choice of $\Lambda$ is physically motivated and indicates the
correction to the $\rho$ mass at the physical point is negligible.
The dot-dashed curve illustrates the correction when $\Lambda^2 = 2$
GeV${}^2$.

   Figure \ref{DRhoMass} indicates that $\Sigma_{\rho\pi\pi}$ is less
than 6\% of the squared $\rho$ mass for $\Lambda^2 < 2$ GeV${}^2$.
This corresponds to a less than 3\% correction to the $\rho$ mass
itself.  The conclusion that may be drawn from this analysis, which
differs from previous considerations of these issues, is that the
predictions of the GF11 group \cite{butler93} are not ``too good to be
true''.  The channel specific nonperturbative correction to the
$\rho$-meson mass may actually be rather small.

   The magnitude of the corrections estimated here is much smaller
than that anticipated in the analysis of Ref. \cite{geiger90} for any
reasonable choice of $\Lambda$.  Geiger and Isgur predict an
unquenching correction to the $\rho$ mass of approximately $-160$ MeV
in contrast to our prediction of a 0 to 25 MeV reduction of the $\rho$
mass.

\section{FULL QCD CALCULATIONS}

   In full QCD simulations, the two-pion induced self-energy
corrections are, of course, already included.  However, the continuum
predictions derived here will differ from those anticipated on the
lattice \cite{degrand91} largely due to the discretization of the
momenta on the lattice and the lattice regularization itself.  This
renders the divergent integral of (\ref{oneloop}) to a finite sum over
a few two-pion states.  In fact, if one hopes to recover the continuum
physics, it will be necessary to first correct the hadron masses
determined at unphysical quark masses to the continuum, infinite
volume limit prior to extrapolating to physical quark masses.

   Figure \ref{gf11} suggests that it may be extremely difficult to
see the effects of virtual two-pion intermediate states in full QCD.
Since a great deal of the integral strength is lost for current
lattice regularization parameters, the correction curve in figure
\ref{gf11} is most likely an optimistically large deviation from the
linear relation.  Even when full QCD calculations reach the current
state of quenched calculations, it is questionable whether one will be
capable of discerning the effects of virtual two-pion states of the
$\rho$ while the $\rho$ is stable.  Of course, once the $\rho$ becomes
unstable and decays to two pions, the lower lying two-pion states will
need to be subtracted from the correlation function prior to
extracting the $\rho$ mass.  This renders an examination of the
$\rho$-meson mass above the two-pion cut nearly impossible.  It should
be mentioned that these arguments support the recent findings of Ref.
\cite{bernard93} where an attempt to observe the effects of virtual
pion states in $\rho$ correlation functions failed.

\begin{figure}
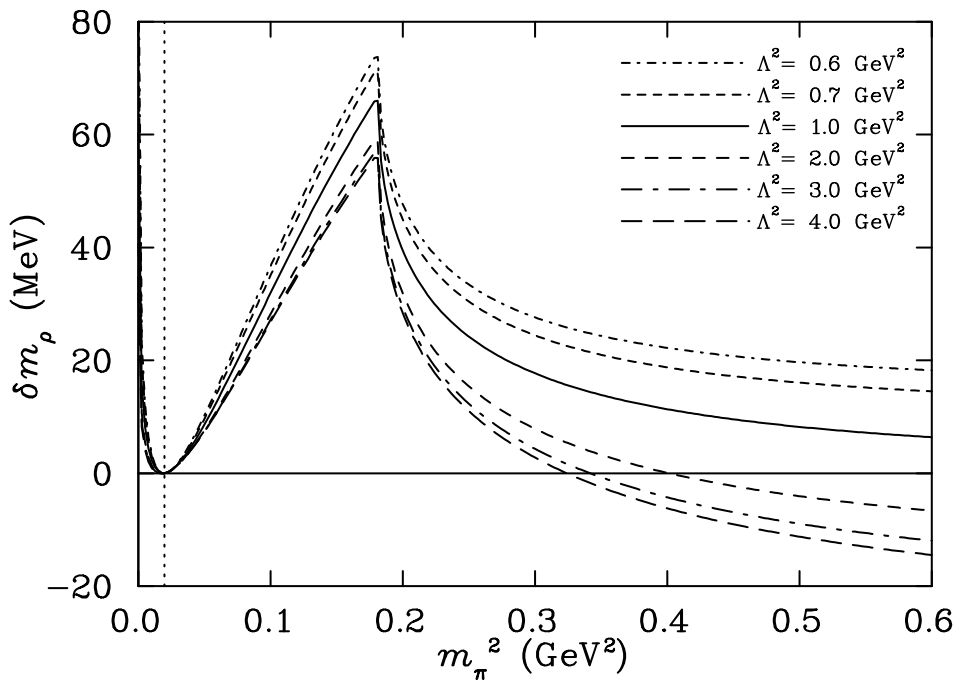

\caption{
The amount to be added to the $\rho$ mass extracted from a
linear extrapolation of full QCD data.  The $x$-axis indicates the
point at which the derivative is determined for the linear
extrapolation.  Values for $\Lambda^2$ are as in figures
\protect\ref{thetacut} and \protect\ref{DRhoMass}.
}
\label{DCorrRhoMass}
\end{figure}

   Ultimately, full QCD calculations will reach the point where
linearly extrapolating the $\rho$ mass to physical quark masses
ignoring nonanalytic effects will introduce a relevant systematic
error.  To this end we present figure \ref{DCorrRhoMass} which
illustrates the relative amount which should be added onto the $\rho$
mass extracted from a linear extrapolation of full QCD data corrected
to the continuum, infinite volume limit \cite{mrhompi}.  We remind the
reader that these effects are being calculated using the dipole form
for the cutoff which we have more or less arbitrarily chosen.  Thus
the curves are illustrative only.  The $x$-axis indicates the point at
which the derivative is determined for the linear extrapolation.
Values for $\Lambda^2$ are as in figures \ref{thetacut} and
\ref{DRhoMass}.  For example, if the effective physical value for
$\Lambda^2$ is 1 GeV${}^2$ and the derivative is determined at
$m_\pi^2 \simeq 0.25$ GeV${}^2$, the $\rho$ mass extracted from a
linear extrapolation should be augmented by approximately 25 MeV prior
to comparing with experiment.

   The size of the corrections to linear extrapolations illustrated in
figure \ref{DCorrRhoMass} are generally smaller than the 70 MeV
addition predicted by Geiger and Isgur.  However, it is possible to
recover a correction to the $\rho$-meson mass extrapolation similar in
magnitude to that of Geiger and Isgur's analysis, provided one
linearly extrapolates from the onset of the cut as in their
investigation.  Of course, it is not practical to extrapolate from the
onset of the cut in actual lattice calculations.  In fact, figure
\ref{DCorrRhoMass} indicates that this is the worst possible place
to attempt an extrapolation to physical quark masses.  The derivative
is more likely to be averaged from a number quark masses corresponding
to squared pion masses of 0.2 to 0.6 GeV${}^2$.  Hence the
corrections are expected to be the order of 10 to 20 MeV and
possibly negligible if the effective physical value for $\Lambda^2$ is
2 GeV${}^2$ or more.

   An important point to mention here is that a simple calculation of
the $\rho$-meson mass in full QCD will not circumvent the problems
associated with its decay to pions.  In fact, as one works hard to
drive down the quark mass, linear extrapolations to the physical point
will increasingly underestimate the $\rho$ mass.  Determination of the
last few percent of the physical $\rho$-meson mass requires additional
information describing the $q^2$ dependence of $\rho\pi\pi$
interactions such that a suitable extrapolation function may be
identified.  Fortunately, these corrections are small and may be
neglected until systematic uncertainties associated with the finite
volume and finite spacings of the lattice are understood and
eliminated.

\section{SUMMARY}

   We have calculated the two-pion induced self-energy correction to
the $\rho$-meson mass in a manner that allows an estimation of the
correction to quenched QCD calculations and an analysis of
extrapolations of full QCD results.

   The analysis of full QCD extrapolations indicates linear
extrapolations of typical $\rho$-meson masses extracted from lattice
correlation functions will underestimate the physical $\rho$ mass by
10 to 20 MeV.  An important point to draw from the analysis is that
as the quark masses become lighter, linear extrapolations to the
physical point will increasingly underestimate the $\rho$-meson mass.

   We estimate the corrections to quenched calculations to be the
order of a few percent and quite possibly negligible (0 to $-25$ MeV).
These results lend credence to the success of quenched QCD in
describing the physical low-lying hadron mass spectrum.

\acknowledgements

\begin{figure}
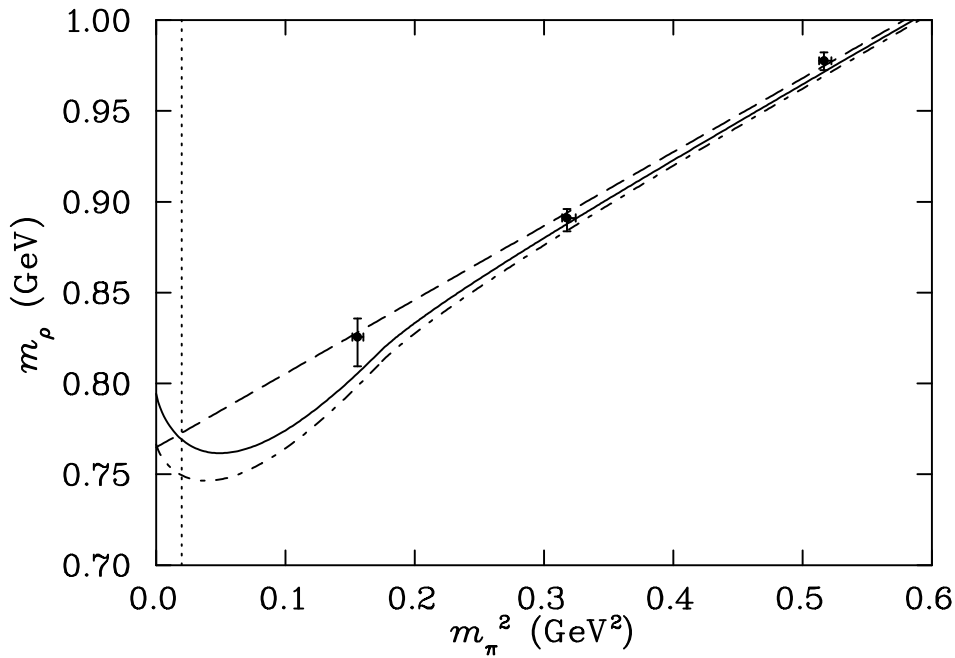

\caption{
$\rho$ and squared pion masses for the three lightest quark masses
used in the quenched QCD analysis of the GF11 group
\protect\cite{butler93} on their largest lattice.  The dashed line
illustrates the linear relationship assumed in extrapolating the
hadron masses to the critical point.  The solid curve displays the
self-energy correction for a regulator mass of $\Lambda^2 = 1$
GeV${}^2$ where the correction to the $\rho$ mass at the physical
point is negligible.  The dot-dash curve corresponds to $\Lambda^2 =
2$ GeV${}^2$.  A lattice scale parameter of 2.73 GeV has been applied
to the otherwise dimensionless lattice results.}
\label{gf11}
\end{figure}

We thank Don Weingarten and Hong Chen for providing us with the
lattice data illustrated in figure \ref{gf11}.  We also thank Manoj
Banerjee for his objective input, and Michael Herrmann for helpful
conversations.  This work is supported in part by the U.S. Department
of Energy under grant DE-FG02-93ER-40762.  T.D.C. acknowledges
additional financial support from the National Science Foundation
though grant PHY-9058487.


\begin{thebibliography}{10}
\bibitem{paramspace}
Ultimately, these uncertainties will be eliminated by directly
calculating the observables in full QCD calculations. However,
explorations of the parameter space for full QCD are in the early
stages. A thorough understanding and elimination of lattice
systematics in full QCD remains a focus of current investigations.
\bibitem{lighteta}
Only the first two terms of the geometric series of pseudoscalar-meson
and glueball states (which can generate the $\eta'$ mass when
geometrically summed) survive in the quenched approximation.
\bibitem{corrlatt}
Ideally, the lattice results should be corrected to the infinite
volume, continuum limit prior to adding the self-energy correction.
\bibitem{mrhompi}
In preparing figure \ref{DCorrRhoMass} we have accounted for the quark
mass (or $m_\pi$) dependence of the $\rho$-meson mass. For
illustrative purposes we set $m_\rho = m_0 + 0.5\, ({\rm GeV}^{-1}) \:
m_\pi^2$ and fixed $m_0$ to reproduce the physical $\rho$ mass at the
physical pion mass.
\end{thebibliography}

\end{document}